# A Predictive Model of Digital Information Engagement: Forecasting User Engagement With English Words by Incorporating Cognitive Biases, Computational Linguistics and Natural Language Processing

Dvir, N.,[1] Friedman, E.,[1] Commuri, S.,[1] Yang, F.[1], Romano, J.[2]

[1]State University of New York at Albany, Albany, New York, NY

[2] Google, New York City, NY

**Corresponding author:**

Nim Dvir

Department of Information Systems and Business Analytics

University at Albany, State University of New York

1400 Washington Avenue, Albany, NY 12222

ndvir@albany.edu

**Declaration of Interest statement:**

This research did not receive any specific grant from funding agencies in the public, commercial, or not-for-profit sectors.

**Abstract**

This study introduces and empirically tests a novel predictive model for digital information engagement (IE) - the READ model, an acronym for the four pivotal attributes of engaging information: Representativeness, Ease-of-use, Affect, and Distribution. Conceptualized within the theoretical framework of Cumulative Prospect Theory, the model integrates key cognitive biases with computational linguistics and natural language processing to develop a multidimensional perspective on information engagement. A rigorous testing protocol was implemented, involving 50 randomly selected pairs of synonymous words (100 words in total) from the WordNet database. These words' engagement levels were evaluated through a large-scale online survey (n = 80,500) to derive empirical IE metrics. The READ attributes for each word were then computed and their predictive efficacy examined. The findings affirm the READ model's robustness, accurately predicting a word's IE level and distinguishing the more engaging word from a pair of synonyms with an 84% accuracy rate. The READ model's potential extends across various domains, including business, education, government, and healthcare, where it could enhance content engagement and inform AI language model development and generative text work. Future research should address the model's scalability and adaptability across different domains and languages, thereby broadening its applicability and efficacy.

# 1 **Introduction**

Information Engagement (IE) denotes the extent to which individuals interact with and process information, a critical concept intersecting several domains, including education, marketing, public policy, and healthcare (Dvir, 2022). It holds the potential to shape attitudes, behaviors, and decisions (Arapakis et al., 2014; Attfield et al., 2011; Dvir, 2020), thus making it a valuable metric for assessing information quality and organizational information-management strategies (O'Brien, 2018). The Advertising Research Foundation (ARF) even declared engagement as the metric of the 21st century for evaluating marketing communication's efficiency and effectiveness (Fulgoni, 2016)

Recent advancements in generative text and AI language models, such as ChatGPT, have presented new avenues to explore IE. These models, powered by Natural Language Processing (NLP) and computational linguistics, can simulate interactive human-like conversations, thereby promising greater insights into IE (Dvir & Gafni, 2019). However, despite this potential, systematic and computational approaches to understanding and improving IE remain scarce (O'Brien & Cairns, 2016).

This study aims to bridge this gap by developing a predictive model leveraging textual data to estimate IE. This model integrates NLP and computational linguistics to analyze IE, while also considering the cognitive biases that shape user engagement with text. Given the rapid advancements in AI and generative language models, understanding the interplay between cognitive biases and user engagement is critical. This research will thus contribute to the existing literature and help optimize user engagement in AI-driven text generation, fostering more personalized and engaging digital interactions.

## 1.1 Objectives

This study's primary objective is to understand the factors that influence IE and to develop a reliable predictive method. Drawing from behavioral economics, we established a multidisciplinary approach to identify IE's predictive attributes, mapping these to quantifiable textual features. We subsequently empirically validated these features as significant predictors.

By doing so, we aim to significantly impact various fields, such as education, marketing, and healthcare, by offering a nuanced understanding of IE that can guide attitudes, behaviors, and decision-making. This research seeks to answer two primary research questions:

**R$_1$: What are the significant predictors of Information Engagement?**

**R$_2$: Can Information Engagement be systematically and predictably determined?**

## 2 Literature review

### 2.1 2.1 The Criticality of Information Engagement

Various studies demonstrate the positive impact of engagement with textual information across different sectors (Dvir, 2020; O'Brien et al., 2020). However, stimulating user engagement with digital content can be challenging due to diverse user motivations (O'Brien et al., 2018; O'Brien & McKay, 2018). User engagement is perceived as an immersive experience requiring cognitive and psychological investment, with information design or the expressive quality of user interface components playing a significant role (Dvir, 2018; Mollen & Wilson, 2010; O'Brien, 2011). Despite the significance of creating engaging content, literature lacks comprehensive guidelines to achieve it (O'Brien & Cairns, 2016).

## 2.2 The User Engagement Scale

O'Brien and Toms developed the User Engagement Scale (UES) to evaluate outcomes of user-system interaction based on attributes of the user and the system (O'Brien et al., 2018; O'Brien & Toms, 2010). The UES measures four self-reported affordances of the information experience: Aesthetic appeal, Focused attention, Perceived usability, and Reward. Aesthetic appeal (AE) pertains to the visual appearance of a computer application interface and its ability to evoke curiosity and interest in users. Focused attention (FA) describes the level of mental concentration resulting from a user's enjoyment and interest during the interaction. Perceived usability (PU) is the user's affective and cognitive responses to a system when using it, such as frustration, and the effort needed to use it. Finally, reward (RW) refers to a user's overall evaluation of the experience of using a system, including perceived success, likelihood of returning, and willingness to recommend the system to others.

The UES has been used to examine UE with various technologies and been adopted by more than 50 international research teams (O'Brien et al., 2018; O'Brien & Cairns, 2016). However, the UES has limitations, such as not emphasizing engagement with information and not focusing on the development of engaging experiences. Self-report metrics measured by the UES may not always be applicable in information environments, so holistic approaches to evaluation are needed. Despite its widespread use, the UES lacks emphasis on engagement with information and on developing engaging experiences, necessitating a more comprehensive evaluation approach.

## 2.3 Computational Approaches to Information Engagement

Several studies have explored computational methodologies to understand engagement, using semi-automated sentiment analysis and NLP techniques. Among them, Berger and Milkman performed semi-automated sentiment analysis to quantify the affectivity and emotionality of *New York Times* articles, while Guerini et al. employed NLP techniques to explore text virality in social networks (Berger & Milkman, 2012; Guerini et al., 2011). Other studies investigating information attributes have used textual analysis and computational linguistics tasks relating to IE to predict the memorability of movie quotes (Cheng et al., 2012) or the comedic value of cartoon captions (Shahaf et al., 2015). The findings of these studies indicated that simple syntax, generality, and distinctiveness are important attributes of engagement.

In relation to specific words, computational linguistics has been used to predict word popularity, word frequency, and the positive affect engendered by words (Turney & Mohammad, 2019). Dvir and Gafni found that in addition to using positive phrasing, information that uses simple, clear language as opposed to complex, technical language creates more incentive to engage with information (Dvir & Gafni, 2018). These aspects—memorability, entertainment value, frequency of words and phrases, and positive affect—are key to understanding IE. Taken together, these findings suggest that information should be expressed as simply, clearly, and inclusively as possible, taking into account the nature of the information being transmitted.

## 2.4 **Summary**

Information Engagement, characterized by sustained user interest, motivation, and attention, is a critical factor in determining the success of a system. However, the literature lacks consensus on the stable, controllable determinants of IE, hindering the systematic application of data analysis, computational linguistics, and NLP to optimize information design and enhance IE. There is a notable dearth of research examining the relationship between content strategy and IE, with most studies focusing on user reactions rather than reasons for engagement or on how to develop engaging content. This study aims to address these gaps by developing a systematic approach to optimizing IE.

## 3  Conceptual Framework and Hypothesis Development

### 3.1  The Underpinning of Cumulative Prospect Theory

This study is grounded in the Cumulative Prospect Theory (CPT), which addresses the role of cognitive biases in user engagement. CPT, a behavioral economics theory, emphasizes the significance of information context and processing in satisfying a user's affective and behavioral information needs (Kahneman & Tversky, 1979; Tversky & Kahneman, 1992). The theory outlines distinct decision-making phases, acknowledging a user's cognitive limitations under risk and uncertainty. It categorizes cognitive processes into two strategies: reason and intuition, each leading to different decision-making approaches. Recognizing the influence of cognitive biases on user engagement is crucial to accurately predicting and enhancing the efficacy of digital information interactions.

### 3.2  The 'READ' Framework of Predictive Attributes

The literature review reveals four predictive dimensions of engaging words—Representativeness, Ease-of-use, Affect, and Distribution (READ). These aspects, connected to a word's perceived usability, cognitive load, sensory appeal, and recognizability, respectively, are derived from empirically validated heuristics predicting decision-making. Each predictor and its operationalization are described below:

#### 3.2.1  Representativeness

Representativeness refers to the degree of resemblance between a new stimulus and a certain standard. Individuals tend to evaluate similarity and organize objects based on their association, leading to increased functionality, familiarity, findability, and fluency (Kahneman & Frederick, 2002). It can be operationalized through semantic relation analysis, evaluating equivalency, hierarchy, and association.

### 3.2.2 Ease-of-use

Ease-of-use pertains to the preference for highly readable texts. Readability and perceptual fluency have been shown to be stimuli and determinants of decision making, perception, and memory (Alter & Oppenheimer, 2008; Tversky & Kahneman, 1974). The ease-of-processing heuristic, which guides and biases judgments about memory, is a mental shortcut that helps individuals understand the world by using information that they find easy to recall and process. It can be operationalized in terms of word length and syllable number, using readability tests like the Flesch–Kincaid reading tests.

### 3.2.3 Affect

Affect refers to the emotional association a word or phrase carries. Affective valence has been shown to influence decision making and ultimately engagement (Batra & Ray, 1986; Commuri & Ekici, 2008; Finucane et al., 2000). The affect heuristic explains why individuals often rely on their emotions when making decisions, which can lead to suboptimal choices. It can be operationalized by performing sentiment analysis, assessing whether a text is positive, negative, or neutral.

### 3.2.4 Distribution

Distribution refers to the frequency, dispersion, and distinctiveness of a piece of information. It relates to well known cognitive biases, such as availability and recognition, According to which, in the process of judgment and decision making individuals tend to rely on information that is easy to recall (Tversky & Kahneman, 1992). When comparing familiar and unfamiliar phrases, users base their judgment on which phrases are more pleasing based on ease of retrieval or recognition (Kahneman & Frederick, 2002). Research has found that frequently read words are read and understood more quickly and easily and are perceived and interpreted more correctly and easily than infrequently read words (Brysbaert et al., 2011; Savin, 1963). In a study of the words used in the Internet Movie Database (IMDB), Dvir and Gafni (2019) found that word frequency impacted IE directly, namely that high-frequency words have a positive impact on IE (Dvir & Gafni, 2019).

It can be operationalized by word frequency, indicating familiarity, cultural association, and readability.

### 3.2.5 The Predictive Model

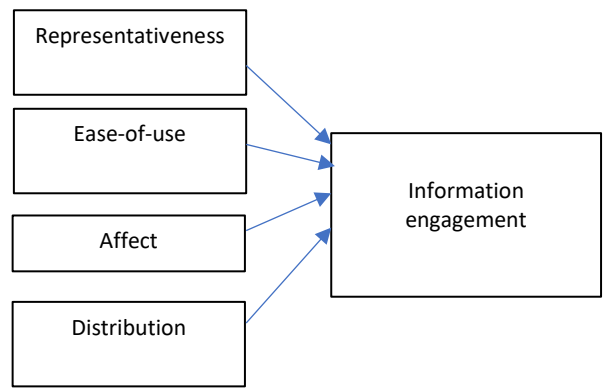

**Figure 1. The READ model**

The READ model, developed in this study, predicts Information Engagement (IE) based on the quantitative assessment of the four textual attributes. It provides a systematic way of determining the probability of a word or text being engaging. The table below summarizes the attributes of the model:

Table 1. Attributes of the READ model

| Attribute | Definition | Factor | Measurement |
|---|---|---|---|
| **Representativeness** | Degree of similarity with a standard | Familiarity | Semantic relation |
| **Ease-of-use** | Complexity and cognitive load | Fluency | Simplicity |
| **Affect** | Emotional association | Feeling | Sentiment analysis |
| **Distribution** | Frequency and recognizability | Availability | Saliency/significance |

Based on the model, we propose the following hypotheses:

*$H_1$: IE with information can be predicted by its representativeness, ease-of-use, affect, and distribution levels.*

*$H_2$: When there are several ways to express the same information, it is possible to predict which one will be more engaging by comparing their representativeness, ease-of-use, affect, and distribution levels.*

## 4 Methods

### 4.1 Study Design

We tested our model by selecting 100 words grouped into 50 synonym sets (synsets) from WordNet, a lexical database of English phrases (Princeton University, 2018). Each word underwent feature extraction based on our model, READ. The model's performance was assessed by predicting each word's Information Engagement (IE), using data from a large-scale survey (n = 80,500). This survey asked participants to evaluate randomly selected words using the User Engagement Scale (UES). Finally, we conducted a multiple linear regression analysis to determine the effectiveness of our model.

The table below lists the 100 words examined, grouped into 50 synonymous pairs (synsets).

**Table 2. Word pairs examined**

| Word₁ | Word₂ | Synset | Definition |
|---|---|---|---|
| **abused** | **maltreated** | abused.a.02 | Subjected to cruel treatment |
| **star** | **maven** | ace.n.03 | Someone who is dazzlingly skilled in any field |
| **quick** | **nimble** | agile.s.01 | Moving quickly and lightly |
| **rich** | **plenteous** | ample.s.02 | Affording an abundant supply |
| **annoying** | **nettlesome** | annoying.s.01 | Causing irritation or annoyance |
| **art** | **prowess** | art.n.03 | A superior skill learned by study, practice, and observation |
| **gone** | **deceased** | asleep.s.03 | Dead |
| **zombie** | **automaton** | automaton.n.01 | Someone who acts or responds in a mechanical or apathetic way |
| **greedy** | **avaricious** | avaricious.s.01 | Immoderately desirous of acquiring something, typically wealth |
| **king** | **magnate** | baron.n.03 | A very wealthy or powerful businessman |
| **mother** | **engender** | beget.v.01 | Make children |
| **bubbling** | **belching** | burp.v.01 | Expel gas from the stomach |
| **fighter** | **belligerent** | combatant.n.01 | Someone who fights or is fighting |
| **computerization** | **cybernation** | computerization.n.01 | The control of processes by computer |
| **cut** | **shortened** | cut.s.03 | With parts removed |
| **lady** | **gentlewoman** | dame.n.02 | A woman of refinement |
| **death** | **demise** | death.n.04 | The time at which life begins to end and continuing until death |
| **going** | **departure** | departure.n.01 | The act of departing |
| **surrogate** | **deputy** | deputy.n.04 | A person appointed to represent or act on behalf of others |
| **find** | **uncovering** | discovery.n.01 | The act of discovering something |
| **dove** | **peacenik** | dove.n.02 | Someone who prefers negotiations to armed conflict in the conduct of foreign relations |
| **done** | **coif** | dress.v.16 | To arrange attractively |
| **enemy** | **opposition** | enemy.n.02 | An armed adversary, especially a member of an opposing military force |
| **foodie** | **epicurean** | epicure.n.01 | A person devoted to refined, sensuous enjoyments, especially good food and drink |
| **exile** | **deportee** | exile.n.02 | A person who is expelled from a home or country by authority |

| Word₁ | Word₂ | Synset | Definition |
|---|---|---|---|
| lush | profuse | exuberant.s.03 | Produced or growing in extreme abundance |
| sticky | mucilaginous | gluey.s.01 | Having the sticky properties of an adhesive |
| hit | smasher | hit.n.03 | A conspicuous success |
| ill | poorly | ill.r.01 | In a poor or improper or unsatisfactory manner; not well |
| now | forthwith | immediately.r.01 | Without delay or hesitation; with no time intervening |
| reciprocation | interchange | interchange.n.02 | Mutual interaction; the activity of reciprocating or exchanging, especially information |
| natural | lifelike | lifelike.s.02 | Free from artificiality |
| just | merely | merely.r.01 | And nothing more |
| motive | need | motivation.n.01 | The psychological feature that arouses an organism to action toward a desired goal; the reason for the action that which gives purpose and direction to behavior |
| being | organism | organism.n.01 | A living thing that has or can develop the ability to act or function independently |
| pale | blanch | pale.v.01 | Turn pale, as if in fear |
| puff | gasp | pant.v.01 | Breathe noisily, as when one is exhausted |
| soul | mortal | person.n.01 | A human being |
| pirate | buccaneer | pirate.n.02 | Someone who robs at sea or plunders the land from the sea without having a commission from any sovereign nation |
| dress | primp | preen.v.03 | Dress or groom with elaborate care |
| rot | putrefaction | putrefaction.n.01 | A state of decay usually accompanied by an offensive odor |
| real | tangible | real.s.04 | Capable of being treated as fact |
| sex | gender | sex.n.04 | The properties that distinguish organisms on the basis of their reproductive roles |
| termination | ending | termination.n.05 | The act of ending something |
| right | veracious | veracious.s.02 | Precisely accurate |
| wash | lave | wash.v.02 | Cleanse one's body with soap and water |
| best | easily | well.r.03 | Indicating high probability; in all likelihood |
| better | well | well.r.11 | In a manner affording benefit or advantage |
| witch | wiccan | wiccan.n.01 | a believer in Wicca |

## 4.2 Instrument

Data collection, including demographic data and survey responses, was done using the QualtricsXM cloud-based survey platform.

## 4.3 Procedure and Measurements

Participants were presented with a word randomly selected from the dataset and were asked to evaluate it based on statements adopted from the UES (O'Brien et al., 2018). The statements were used to assess engagement based on sensory appeal, ability to stimulate attention, perceived usability, and reward. The statements were presented in random order, with 4 positively formulated (e.g., "This word is easy to understand") and 4 negatively formulated (e.g., "This word is difficult to understand"). Participants responded using a 5-point scale, with responses to negative statements reverse-coded to maintain a unified scale.

Table 3. Statements for IE Evaluation Adopted From the UES

| Code | Statement |
|------|-----------|
| EA | This word appealed to my senses. |
| EA-n | This word is not engaging. |
| FA | This word drew my attention. |
| FA-n | I wasn't focused while reading this word. |
| PU | This word was easy to understand. |
| PU-n | This word was difficult to understand. |
| RW | The experience of reading this word was rewarding. |
| RW-n | The experience of reading this word is not worthwhile. |

Testing for scale reliability yielded a Cronbach's alpha of 0.888. The mean UES of the words was 2.44 (SD = .877). We have made the UES scores of all the words available in a public data repository (Dvir, 2023a).

### 4.4 **Participants**

Participants were undergraduate students recruited from a large research university in the United States. The recruitment method and use of a survey were reviewed and approved by the University at Albany Institutional Review Board (IRB Study No. 22X113).

The mean age of the participants was 22.1 (SD = 1.388), with 75.3% of responses made by the age group 17–22, and 41.2% were females and 58.8% males. More than 95.0% reported that their English proficiency was at least very good, with 79.0% identifying as native speakers. The majority of the observations were submitted through a laptop device (70.1%), followed by mobile devices (15.6%) and desktops (14.2%). These percentages may be attributed to the survey being easier to read and complete on a large screen, as was reported by participants in early stages of the survey design.

We used Qualtrics software to verify that the survey was only completed once, therefore controlling for unique participants. Our system randomly presented each word of the dataset 805 times to randomly selected participants, thereby controlling for participant characteristics. Overall, we collected 80,500 observations.

By balancing both known and unknown predictive factors, our survey design aimed to reduce biases, such as selection bias and allocation bias, to the greatest extent possible. Chi-square analysis of the goodness-of-fit of the samples revealed that the characteristic composition for each word sample was comparable to that of the overall population. Pearson chi-square analysis between demographic groups and between word samples revealed no significant differences in the composition of the demographic groups who responded to each of the words displayed.

## 4.5  Feature Extraction

For each word we extract textual features based on the READ model. We used the Natural Language Toolkit (NLTK) and the Python programming language to transform each word into a numerical representation for Natural Language Processing (NLP) classification. The quantitative data obtained includes measures related to the word's representativeness, ease-of-use, affect, and distribution.

### 4.5.1  Representativeness

Operationalizing representativeness requires conducting semantic relation analysis to calculate equivalency, hierarchy, and association. We used the following predictors:

1. **Definitions:** A sense or definition represents one aspect of a word's meaning. We determined polysemy, the capacity for a word to have multiple meanings, using WordNet, a lexical database of English phrases grouped into sets of cognitive synonyms. The mean definition score for all words was 7.85 (SD = 11)

2. **Hypernyms:** A hypernym is a superordinate term that encompasses more specific words. For example, the word "color" is a hypernym of "red." We used WordNet to analyze noun hierarchies, ultimately leading up to the root node (entity). The mean number of hypernyms for each word was 4.75 (SD = 7.5).

3. **Hyponyms:** A hyponym is a subordinate term that falls under one or more broader terms. For example, the word "banana" is a hyponym (more specific concept) for the word "fruit." We used WordNet to analyze noun hierarchies, ultimately leading up to the root node (entity). The mean number of hypernyms for each word was 26.41 (SD =70.015).

### 4.5.2  Ease-of-use features

We used the following measures to evaluate the ease-of-use of a word:

1.  **Length:** Measured by character count using the NLTK, the mean length of words in the list was 6.58 (SD = 2.62)
2.  **Syllable count:** Measured manually, the mean syllable count for the list was 2.17 (SD = 1.14).
3.  **Flesch Reading Ease Score:** Calculated using Textstat library, this score provides a score between 1 and 100, with higher scores indicating higher readability. The mean score for the list was 36.6 (SD = 89.98).

### 4.5.3 Affect Features

We used sentiment analysis to assess whether a word expressed a positive, negative, or neutral sentiment. For this we used SentiWordNet 3.0, a lexical database based on WordNet, to assign an overall sentiment score to each word (Baccianella et al., 2010) based on three sentiment scores were calculated:

1.  **Positivity:** We based the positivity score on all the synset scores associated with a word, choosing the maximum positivity score. The score ranged from 0 (minimum positivity) to 1 (maximum positivity). The mean score for all words was 0.22 (SD = .267).
2.  **Negativity:** We based the negativity scores on all the synset scores associated with a word, choosing the maximum score. The score ranged from 0 (minimum negativity) to 1 (maximum negativity). The mean score for all words was 0.197 (SD = 0.26).

### 4.5.4 Distribution measures

1.  **Frequency:** We operationalized distribution as the frequency of words using WordFreq, a Python library that identifies word frequency using data from the Exquisite Corpus, which compiles data from various domains (Speer et al., 2018). The domains include encyclopedic text from Wikipedia, subtitles from OPUS OpenSubtitles, news from NewsCrawl and GlobalVoices, books from Google Books Ngrams, and short-form and long-form social media text from Twitter and Reddit, respectively.

We made the original Python program created for feature extraction available through an open-source repository [1]. We have made the full dataset for the read scores of the 100 words that we evaluated available in an open-access repository (Dvir, 2023b).

### 4.6 Prediction and Evaluation

We performed multiple linear regression analysis to examine the relationship between predictor variables (representativeness, ease-of-use, affect, and distribution) and the outcome variable (mean UES score for each word). By such means, we measured the model's accuracy by comparing its predictions with IE values obtained in a user study.

## 5 Results

***$H_1$: IE with information can be predicted by its representativeness, ease-of-use, affect, and distribution levels.***

For hypothesis $H_1$, we conducted multiple linear regressions, taking the User Engagement Score (UES) as the dependent variable, and the words' textual features as the independent variables. While all variables significantly contributed to predicting the UES, the number of definitions was an exception. The model was statistically significant (adjusted $R^2 = 0.539$, F = 9081.442, $p < 0.001$), suggesting it explained a considerable portion of the UES variance. However, we note that $R^2$ alone should not be used to evaluate the model's performance, especially for future or unseen data.

---

[1] https://github.com/nimdvir/READ-model/find/main

## Table 4. Regression Results of UES Prediction Based on READ Features

| Variable | B | SE B | β | t | p |
|---|---|---|---|---|---|
| **(Constant)** | 2.297 | 0.005 | | 443.730 | 0.000 |
| **Definitions-synsets** | 0.000 | 0.000 | -0.003 | -0.704 | 0.481 |
| **Hypernyms** | -0.005 | 0.000 | -0.185 | -40.455 | 0.000 |
| **Hyponyms** | 0.000 | 0.000 | 0.034 | 12.669 | 0.000 |
| **Positive** | 0.126 | 0.005 | 0.161 | 23.911 | 0.000 |
| **Negative** | -0.113 | 0.005 | -0.143 | -24.326 | 0.000 |
| **Length** | -0.019 | 0.001 | -0.241 | -36.046 | 0.000 |
| **Flesch_reading_ease** | -0.001 | 0.000 | -0.328 | -51.952 | 0.000 |
| **Syllable** | -0.041 | 0.001 | -0.223 | -32.292 | 0.000 |
| **Frequency** | -175.563 | 1.842 | -0.302 | -95.297 | 0.000 |

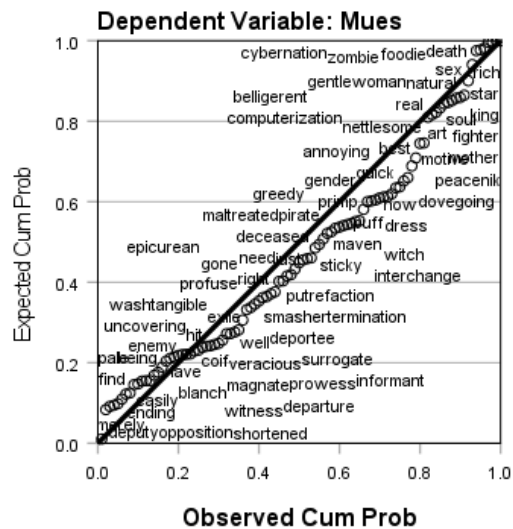

**Figure 2. Normal P-P Plot of Regression Standardized Residual**

***H₂: When there are several ways to express the same information, it is possible to predict which one will be more engaging by comparing their representativeness, ease-of-use, affect, and distribution levels.***

To test hypothesis H₂, we conducted pairwise comparisons of synonyms in 50 synsets. The aim was to determine whether the choice of word impacts the IE. Our results revealed significant differences in the UES scores between the synonyms in 43 out of 50 synsets ($P < 0.05$). The differences in mean UES scores are graphically represented in the figure below.

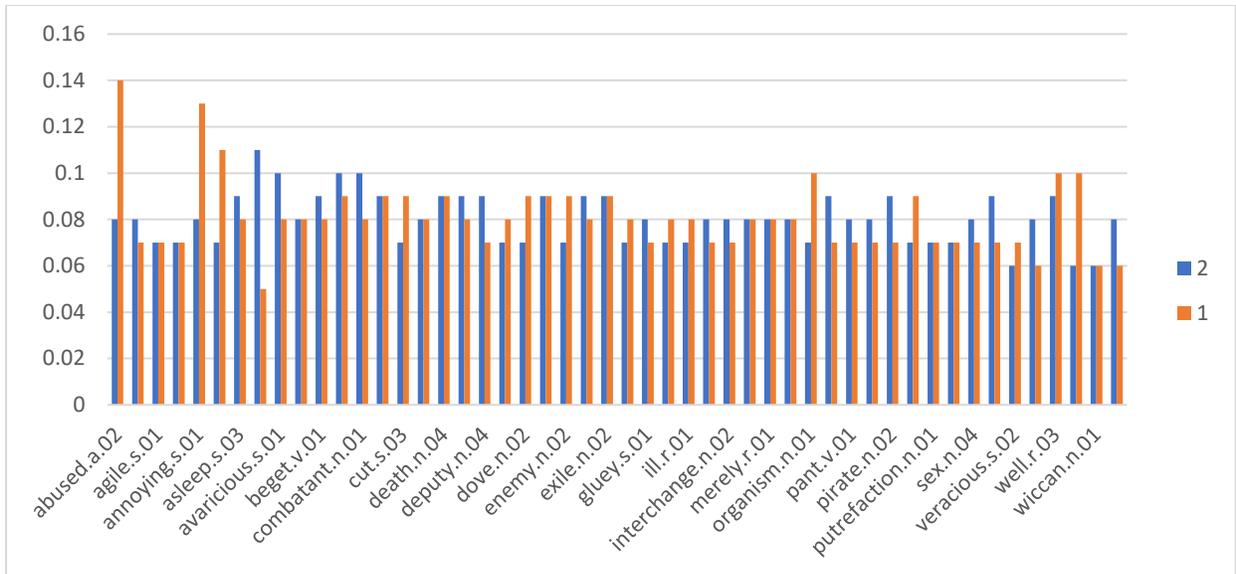

**Figure 3. Difference in mean UES scores between synsets**

Further, we used our UES predictions to determine which word in each synset could potentially have a higher IE score. The model correctly predicted 42 out of the 50 synsets, yielding an accuracy rate of 84%. Table 7 compares the observed and predicted UES scores for each synset.

**Table 5. Observed and Predicted UES Scores for 50 Synsets**

| Synset | Word₁ | UES₁ | Predicted₁ | Word₂ | UES₂ | Predicted ₂ | Higher observed UES score | Higher predicted UES score |
|---|---|---|---|---|---|---|---|---|
| abused.a.02 | abused | 2.31 | 2.41 | maltreated | 2.23 | 2.22 | abused | abused |
| ace.n.03 | star | 2.83 | 2.61 | maven | 2.19 | 2.18 | star | star |
| agile.s.01 | quick | 2.66 | 2.61 | nimble | 2.35 | 2.46 | quick | quick |
| ample.s.02 | rich | 3.07 | 2.67 | plenteous | 2.21 | 2.17 | rich | rich |
| annoying.s.01 | annoying | 2.47 | 2.44 | nettlesome | 2.20 | 2.10 | annoying | annoying |
| art.n.03 | art | 2.82 | 2.69 | prowess | 2.35 | 2.48 | art | art |
| asleep.s.03 | gone | 2.39 | 2.44 | deceased | 2.37 | 2.38 | gone | gone |
| automaton.n.01 | zombie | 2.91 | 2.46 | automaton | 2.35 | 2.31 | zombie | zombie |
| avaricious.s.01 | greedy | 2.47 | 2.45 | avaricious | 2.20 | 2.14 | greedy | greedy |

| Synset | Word₁ | UES₁ | Predicted₁ | Word₂ | UES₂ | Predicted₂ | Higher observed UES score | Higher predicted UES score |
|---|---|---|---|---|---|---|---|---|
| baron.n.03 | king | 2.89 | 2.60 | magnate | 2.22 | 2.37 | king | king |
| beget.v.01 | mother | 2.73 | 2.58 | engender | 2.21 | 2.32 | mother | mother |
| burp.v.01 | bubbling | 2.45 | 2.43 | belching | 2.24 | 2.27 | bubbling | bubbling |
| combatant.n.01 | fighter | 2.62 | 2.49 | belligerent | 2.43 | 2.28 | fighter | fighter |
| computerization.n.01 | computerization | 2.35 | 2.23 | cybernation | 2.31 | 2.03 | computerization | computerization |
| cut.s.03 | cut | 2.46 | 2.44 | shortened | 2.20 | 2.37 | cut | cut |
| dame.n.02 | lady | 2.66 | 2.63 | gentlewoman | 2.44 | 2.26 | lady | lady |
| death.n.04 | death | 2.96 | 2.59 | demise | 2.31 | 2.39 | death | death |
| departure.n.01 | going | 2.45 | 2.39 | departure | 2.31 | 2.46 | going | departure |
| deputy.n.04 | surrogate | 2.30 | 2.40 | deputy | 2.28 | 2.47 | surrogate | deputy |
| discovery.n.01 | find | 2.45 | 2.60 | uncovering | 2.24 | 2.34 | find | find |
| dove.n.02 | dove | 2.54 | 2.49 | peacenik | 2.19 | 2.13 | dove | dove |
| dress.v.16 | done | 2.54 | 2.57 | coif | 2.20 | 2.30 | done | done |
| enemy.n.02 | enemy | 2.48 | 2.60 | opposition | 2.31 | 2.49 | enemy | enemy |
| epicure.n.01 | foodie | 2.70 | 2.43 | epicurean | 2.22 | 2.28 | foodie | foodie |
| exile.n.02 | exile | 2.41 | 2.49 | deportee | 2.18 | 2.24 | exile | exile |
| exuberant.s.03 | lush | 2.50 | 2.48 | profuse | 2.24 | 2.30 | lush | lush |
| gluey.s.01 | sticky | 2.37 | 2.38 | mucilaginous | 2.18 | 2.03 | sticky | sticky |
| hit.n.03 | hit | 2.45 | 2.56 | smasher | 2.32 | 2.37 | hit | hit |
| ill.r.01 | ill | 2.51 | 2.61 | poorly | 2.33 | 2.47 | ill | ill |
| immediately.r.01 | now | 2.54 | 2.50 | forthwith | 2.19 | 2.31 | now | now |

| Synset | Word₁ | UES₁ | Predicted₁ | Word₂ | UES₂ | Predicted ₂ | Higher observed UES score | Higher predicted UES score |
|---|---|---|---|---|---|---|---|---|
| interchange.n.02 | reciprocation | 2.38 | 2.24 | interchange | 2.30 | 2.34 | reciprocation | interchange |
| lifelike.s.02 | natural | 2.76 | 2.61 | lifelike | 2.42 | 2.44 | natural | natural |
| merely.r.01 | just | 2.44 | 2.45 | merely | 2.22 | 2.54 | just | merely |
| motivation.n.01 | motive | 2.57 | 2.48 | need | 2.54 | 2.57 | motive | need |
| organism.n.01 | being | 2.47 | 2.60 | organism | 2.41 | 2.43 | being | being |
| pale.v.01 | pale | 2.38 | 2.52 | blanch | 2.16 | 2.32 | pale | pale |
| pant.v.01 | puff | 2.46 | 2.42 | gasp | 2.37 | 2.44 | puff | gasp |
| person.n.01 | soul | 2.80 | 2.66 | mortal | 2.54 | 2.53 | soul | soul |
| pirate.n.02 | pirate | 2.48 | 2.48 | buccaneer | 2.33 | 2.29 | pirate | pirate |
| preen.v.03 | dress | 2.53 | 2.51 | primp | 2.23 | 2.22 | dress | dress |
| putrefaction.n.01 | rot | 2.40 | 2.45 | putrefaction | 2.21 | 2.24 | rot | rot |
| real.s.04 | real | 2.77 | 2.64 | tangible | 2.47 | 2.57 | real | real |
| sex.n.04 | sex | 3.11 | 2.67 | gender | 2.59 | 2.56 | sex | sex |
| termination.n.05 | termination | 2.36 | 2.40 | ending | 2.34 | 2.53 | termination | ending |
| veracious.s.02 | right | 2.59 | 2.64 | veracious | 2.22 | 2.33 | right | right |
| wash.v.02 | wash | 2.47 | 2.56 | lave | 2.21 | 2.32 | wash | wash |
| well.r.03 | best | 2.79 | 2.71 | easily | 2.52 | 2.70 | best | best |
| well.r.11 | better | 2.61 | 2.71 | well | 2.58 | 2.67 | better | better |
| wiccan.n.01 | witch | 2.46 | 2.46 | wiccan | 2.25 | 2.35 | witch | witch |

In summary, our findings supported both hypotheses. They highlighted the role of textual features in predicting a word's IE and demonstrated that comparing these features can predict which expression of the same information might be more engaging.

# 6  Discussion

This study's findings present robust evidence supporting our two hypotheses and offer answers to the research questions posed.

**R₁: What are the significant predictors of Information Engagement (IE)?**

Four primary attributes of information - representativeness, ease-of-use, affect, and distribution - were identified as significant predictors of user engagement.

**Representativeness** is the degree to which a word encapsulates the overall theme of the information. Our data suggests that words with higher representativeness tend to attract more engagement, implying that content creators should use words that best represent their central message for maximum audience engagement.

**Ease-of-use** denotes the simplicity and accessibility of words. Users showed a preference for words that are simple and easy to comprehend, process, and remember, owing to cognitive fluency, which favors information requiring less mental effort. Thus, using simpler, more familiar words could enhance user engagement.

**Affect** refers to a word's emotional impact. Our study indicated that words with stronger emotional connotations, both positive and negative, fostered higher levels of engagement. This underlines the value of emotionally charged language in content creation.

**Distribution** pertains to the frequency of word occurrence. Words appearing more frequently tend to receive higher user engagement, emphasizing the need for strategic word placement to sustain user attention and enhance engagement.

**R₂: Can IE be systematically and automatically predicted?**

To answer this question, we employed a large-scale user survey and a computational model. Participants interacted with various words and assigned their perceived engagement scores. This user-generated data enhanced our model's accuracy.

We developed a Python program to estimate a word's representativeness, ease-of-use, affect, and distribution by analyzing its textual properties. When tested, our model demonstrated a strong correlation between predicted engagement levels and actual perception scores, indicating its ability to accurately predict word engagement.

Our model also proved effective in predicting which synonym would be more engaging for the same information, systematically determining the most engaging choice in 82% of the cases. These results strongly suggest that IE can be systematically and automatically predicted using our model.

This study bridges two significant research gaps regarding the relationship between IE and textual data—how users react to information and the optimal means of developing engaging information. These findings provide a foundation for future research to further refine the methods of creating engaging information.

## 7 Conclusions

Our study robustly supports the READ model's efficacy as a novel, systematic approach for predicting and enhancing Information Engagement (IE), offering valuable insights for crafting more effective communication strategies tailored to diverse audiences and contexts. Utilizing the knowledge from our study, content creators can refine their language choices to boost user engagement and comprehension, thereby augmenting their overall communication effectiveness.

The next research trajectory should focus on establishing more reliable and reusable metrics for measuring engagement. This would aid in the precise specification of engaging language and enable a quantitative evaluation of IE. A better understanding of engaging language attributes can pave the way for improved and optimized textual data across domains where optimal wording is paramount, such as business, non-profit work, government, and education.

Engagement serves as a valuable framework for understanding, measuring, and enhancing interactions with information. Language can act as a pragmatic toolkit to influence human behavior, with words serving as vessels to carry information, some more potent than others. Future research could leverage this study's findings to create an IE classifier for classifying words and quantitatively measuring IE.

The potential of this classifier to quantify a word's "stickiness"—its memorability and engagement factor— or to create an engagement spectrum based on the identified IE factors is promising. Constructing an automated method for effective communication could revolutionize the way messages are conveyed and received.

Our results hold significant implications for future language-processing tools and techniques aimed at enhancing user engagement. Subsequent research should consider additional factors that may influence information engagement and explore methods to refine the model's predictive accuracy and applicability across different contexts and languages.

Despite its contributions, this study had limitations that might constrain the generalizability of our findings. The primary limitation was the exclusive examination of undergraduate students at a single university. Future studies should extend the investigation to other populations.

Additionally, our use of snowball sampling as the recruitment method, a non-random sampling technique prone to sampling bias, may limit representation. To counter this, future studies should employ different methodologies and techniques to validate and replicate our findings. If successful, they will strengthen the READ model's standing as a powerful tool for predicting and enhancing IE with textual information.

Dvir, N., & Gafni, R. (2018). When less is more: Empirical study of the relation between consumer behavior and information sharing on commercial landing pages. *Informing Science: The International Journal of an Emerging Transdiscipline*, *21*, 019--039. https://doi.org/10.28945/4015

Dvir, N., & Gafni, R. (2019). Systematic improvement of user engagement with academic titles using computational linguistics. *ArXiv Preprint ArXiv:1906.09569*.

Finucane, M. L., Alhakami, A., Slovic, P., & Johnson, S. M. (2000). The affect heuristic in judgments of risks and benefits. *Journal of Behavioral Decision Making*, *13*(1), 1–17.

Fulgoni, G. M. (2016). In the Digital World, Not Everything That Can Be Measured Matters: How to Distinguish "Valuable" from "Nice to Know" Among Measures of Consumer Engagement. *Journal of Advertising Research*, *56*(1), 9–13. https://doi.org/10.2501/JAR-2016-008

Guerini, M., Strapparava, C., & Özbal, G. (2011). Exploring Text Virality in Social Networks. *ICWSM*. http://www.aaai.org/ocs/index.php/ICWSM/ICWSM11/paper/viewFile/2820/3235-accessdate=27

Kahneman, D., & Frederick, S. (2002). Representativeness Revisited: Attribute Substitution in Intuitive Judgment. In T. Gilovich, D. Griffin, & D. Kahneman (Eds.), *Heuristics and Biases* (1st ed., pp. 49–81). Cambridge University Press. https://doi.org/10.1017/CBO9780511808098.004

Kahneman, D., & Tversky, A. (1979). Prospect Theory: An Analysis of Decision under Risk. *Econometrica*, *47*(2), 263–292.
28

Mollen, A., & Wilson, H. (2010). Engagement, telepresence and interactivity in online consumer experience: Reconciling scholastic and managerial perspectives. *Journal of Business Research*, *63*(9–10), 919–925. https://doi.org/10.1016/j.jbusres.2009.05.014

O'Brien, H. L. (2011). Exploring user engagement in online news interactions. *Proceedings of the American Society for Information Science and Technology*, *48*(1), 1–10.

O'Brien, H. L. (2018). A Holistic Approach to Measuring User Engagement. In M. Filimowicz & V. Tzankova (Eds.), *New Directions in Third Wave Human-Computer Interaction: Volume 2—Methodologies* (pp. 81–102). Springer International Publishing. https://doi.org/10.1007/978-3-319-73374-6_6

O'Brien, H. L., Arguello, J., & Capra, R. (2020). An empirical study of interest, task complexity, and search behaviour on user engagement. *Information Processing & Management*, *57*(3), 102226. https://doi.org/10.1016/j.ipm.2020.102226

O'Brien, H. L., & Cairns, P. (2016). *Why engagement matters: Cross-disciplinary perspectives of user engagement in digital media*. Springer. https://doi.org/10.1007/978-3-319-27446-1

O'Brien, H. L., Cairns, P., & Hall, M. (2018). A practical approach to measuring user engagement with the refined user engagement scale (UES) and new UES short form. *International Journal of Human-Computer Studies*, *112*, 28–39. https://doi.org/10.1016/j.ijhcs.2018.01.004

O'Brien, H. L., & McKay, J. (2018). Modeling Antecedents of User Engagement. In *The Handbook of Communication Engagement* (pp. 73–88). Wiley-Blackwell. https://doi.org/10.1002/9781119167600.ch6
29